\pdfoutput=1

\documentclass[
    ,final            
  ,numberedheadings 
  ]
  {aipproc}

\layoutstyle{8x11single}
\usepackage{upgreek}
\usepackage{amsmath}
\usepackage{amssymb}

\newcommand {\delm}{$\Delta m^{2}$}

\newcommand{\numu}{\mbox{$\nu_{\mu}$}}                   
\newcommand{\nue}{\mbox{$\nu_{e}$}}                      

\begin{document}

\title{First MINOS+ Data and New Results from MINOS}


\author{Alexandre B. Sousa\\ {\normalsize (for the MINOS and MINOS+ Collaborations)}}{
  address={Department of Physics, University of Cincinnati, Cincinnati, Ohio 45221, USA}
}

\begin{abstract}
Following a 7-year run with the MINOS experiment from 2005 to 2012, the MINOS+ long-baseline neutrino experiment started operations in September 2013. Utilizing the MINOS Near and Far Detectors at Fermilab and northern Minnesota, respectively, and measuring the new medium-energy NuMI neutrino beam over baseline of 735~km, MINOS+ has accumulated an exposure of $1.7\times10^{20}$~protons-on-target. In this paper, we present preliminary results from the MINOS+ run. In addition, we report on improved results from a three-neutrino flavor analysis of the full $14.1\times10^{20}$ MINOS-era beam sample, and the combined full MINOS and MINOS+ atmospheric neutrino samples, corresponding to an exposure of 48.7~kiloton-years increased by 28\% with respect to previously published MINOS measurements. The analysis combines the measurements of muon neutrino disappearance and appearance into electron neutrinos in the beam and atmospheric samples. Furthermore, we present results from a new search for sterile neutrinos in the complete MINOS NuMI sample. A 3+1 neutrino mixing model with one sterile neutrino is assumed. Neutrino mixing at the ND baseline ($\sim 1$~km), relevant for values of the additional mass eigenstate $m_4\gtrsim$1~eV, is included for the first time in the analysis of Far and Near Detector data from a long-baseline neutrino experiment. Finally, new limits on non-standard neutrino interactions are obtained from measurements of electron neutrino appearance in the full MINOS beam sample. 
\end{abstract}

\maketitle


\section{Introduction}

The {\bf M}ain {\bf I}njector ${\bf N}$eutrino {\bf O}scillation {\bf S}earch (MINOS) experiment is a long-baseline neutrino oscillation experiment using the  {\bf N}e{\bf u}trinos at the {\bf M}ain {\bf I}njector (NuMI) neutrino beam and two detectors  to make precise measurements of neutrino oscillation parameters over a distance of $735\,$km. It aims to determine to which extent muon neutrinos oscillate into tau and electron neutrinos  or potentially sterile neutrinos, as well as investigate exotic scenarios such as non-standard neutrino interactions. The Near Detector (ND), located 1.04~km downstream of the neutrino production target measures the flux, cross sections , and flavor composition of the predominantly muon-neutrino beam. The  Far Detector (FD), located 735~km from the production target, makes the same beam measurements after neutrino oscillations may have occurred. By using two detectors, MINOS achieves extensive cancellation of systematic uncertainties related to the neutrino beam flux and neutrino interaction cross sections.
MINOS makes measurements of muon neutrino and antineutrino disappearance using both beam and atmospheric samples. These results are now combined with searches for subdominant mixing between muon and electron neutrino flavors, which is sensitive to the value of $\theta_{13}$, the octant of $\theta_{23}$, the neutrino mass hierarchy, and the CP-violating phase $\delta_{CP}$. These measurements also afford MINOS sensitivity to more exotic scenarios, including the admixture of additional sterile neutrino flavors, or the existence of new beyond the standard model neutrino interactions. MINOS took data for 7 years, during which it accumulated a total exposure of $14.1\times10^{20}$~protons-on-target after data quality criteria and detector live time are applied. 
Since September 2013, the MINOS experiment has become MINOS+, operating with the same detectors with upgraded electronics and the NuMI beam updated to the running conditions to be used during the NO$\upnu$A era. The new medium-energy beam spectrum peaks at an energy of 7~GeV for the on-axis location of MINOS+, higher than the 3~GeV low-energy beam used by MINOS, and therefore further away from the 1.6 GeV minimum of the muon neutrino survival probability at 735 km. However, the increased beam flux will provide $\sim4000$ \numu~charged-current (CC) neutrino interactions at the FD for the nominal exposure of $6\times10^{20}$\,protons-on-target/year, and the higher beam energy will enable high-precision searches for deviations from the standard three-flavor oscillation picture. As the only wide-band beam long-baseline neutrino experiment operating in this decade, MINOS+ provides opportunities to unravel new neutrino surprises.

\subsection{The NuMI Beam and the MINOS+ Detectors}
The NuMI beam is produced by collisions with a graphite target of protons accelerated to 120 GeV at Fermilab's Main Injector. The secondary products of these collisions, pions and kaons, are focused by two parabolic magnetic horns and eventually decay into muons and neutrinos inside a 675\,m long decay pipe filled with helium. The muons are absorbed in the rock and neutrinos continue towards the 1\,kton ND, 1\,km downstream of the target, and beyond, towards the 5.4\,kton FD.

\begin{figure}[h!]
	\centering
  		\includegraphics[width=0.48\textwidth]{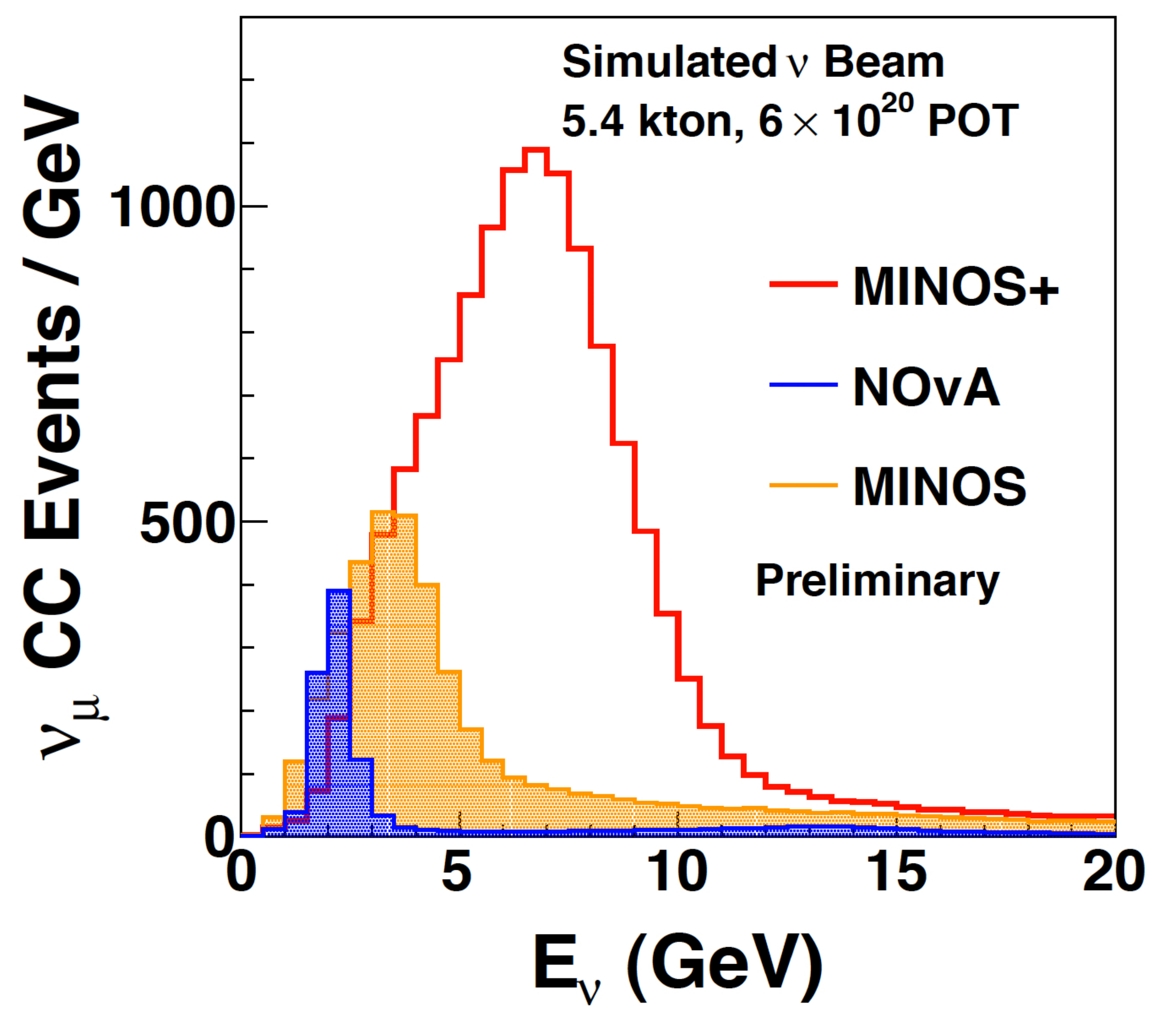}

	\caption{The NuMI neutrino energy spectrum for the MINOS+ medium-energy tune, shown as the red solid line. The NO$\upnu$A~spectrum shown in blue is obtained with the same tune at a 14\,mrad offset from the beam axis (MINOS+ is on-axis). For comparison, the spectrum corresponding to the NuMI low-energy tune used by MINOS is shown as the gold histogram.}
  \label{fig:spectrum}
\end{figure}

For most of the MINOS data taking, the magnetic horns focused positive pions and kaons, resulting in a beam composition measured at the ND of 91.7\% \numu, 7.0\% $\overline{\nu}_\mu$ from decays of very forward negative pions, and  a 1.3\%~($\nu_e~+~\overline{\nu}_e$) fraction primarily from kaon and muon decays.  By inverting the current in the magnetic horns, thus preferentially focusing negative pions, an antineutrino-enhanced beam composition is obtained, with 39.9\% $\overline{\nu}_\mu$, 58.1\% \numu, and 2.0\%~($\nu_e~+~\overline{\nu}_e$). However, the neutrinos in this configuration result from higher-energy forward positive pions that go through the horns undisturbed, so they are preferentially distributed at higher energies than the first minimum of the three-flavor survival probability.
In preparation for operations of the NO$\upnu$A experiment, the NuMI beam line was updated during a year-long shutdown, and resumed operations in September 2013. In the first year of operation, NuMI ran predominately in neutrino mode, with an average power of 280\,kW, corresponding to $2.4\times10^{20}$\,proton collisions in the target for every 10\,$\mu$s beam spill. When ongoing improvements to the Booster ring are completed in 2016, NuMI is expected to attain its nominal beam power of 700\,kW and intensity of $6\times10^{20}$\,protons-on-target/year. The medium-energy and low-energy NuMI beam spectra are shown in Fig.~\ref{fig:spectrum}.

The MINOS+ Near and Far Detectors are planar steel-scintillator tracking calorimeters with a total mass of 1\,kton and 5.4\,kton, respectively. Each detector plane is composed of a 2.54 cm-thick steel layer and a 1-cm thick plastic scintillator layer. The distance between consecutive planes is 5.96\,cm for both detectors. The plastic scintillator layer consists of 4.1-cm wide strips. The strips are oriented in orthogonal directions for successive planes to enable three-dimensional reconstruction of neutrino interactions. The scintillation light is captured by a wavelength shifting fibre embedded along the length of the strips, and read out by multi-anode Hamamatsu photomultiplier tubes. Both detectors are magnetized with a toroidal field with an average value of 1.2\,T in the steel, allowing for the measurement of both charge and momentum of muon tracks from analysis of the track curvature in the field. The FD is installed 705\,m underground in the Soudan Underground Laboratory and is equipped with a veto shield assembled from plastic scintillator strips for rejection of cosmic rays, allowing for studies of atmospheric neutrino data~\cite{ref:minosatmos}.  

\section{Three-flavor combined analysis of \numu~disappearance and \numu$\rightarrow\nu_e$~appearance}\label{sec:threeflavor}
\begin{figure}[h!]
	\centering
  		\includegraphics[width=0.7\textwidth]{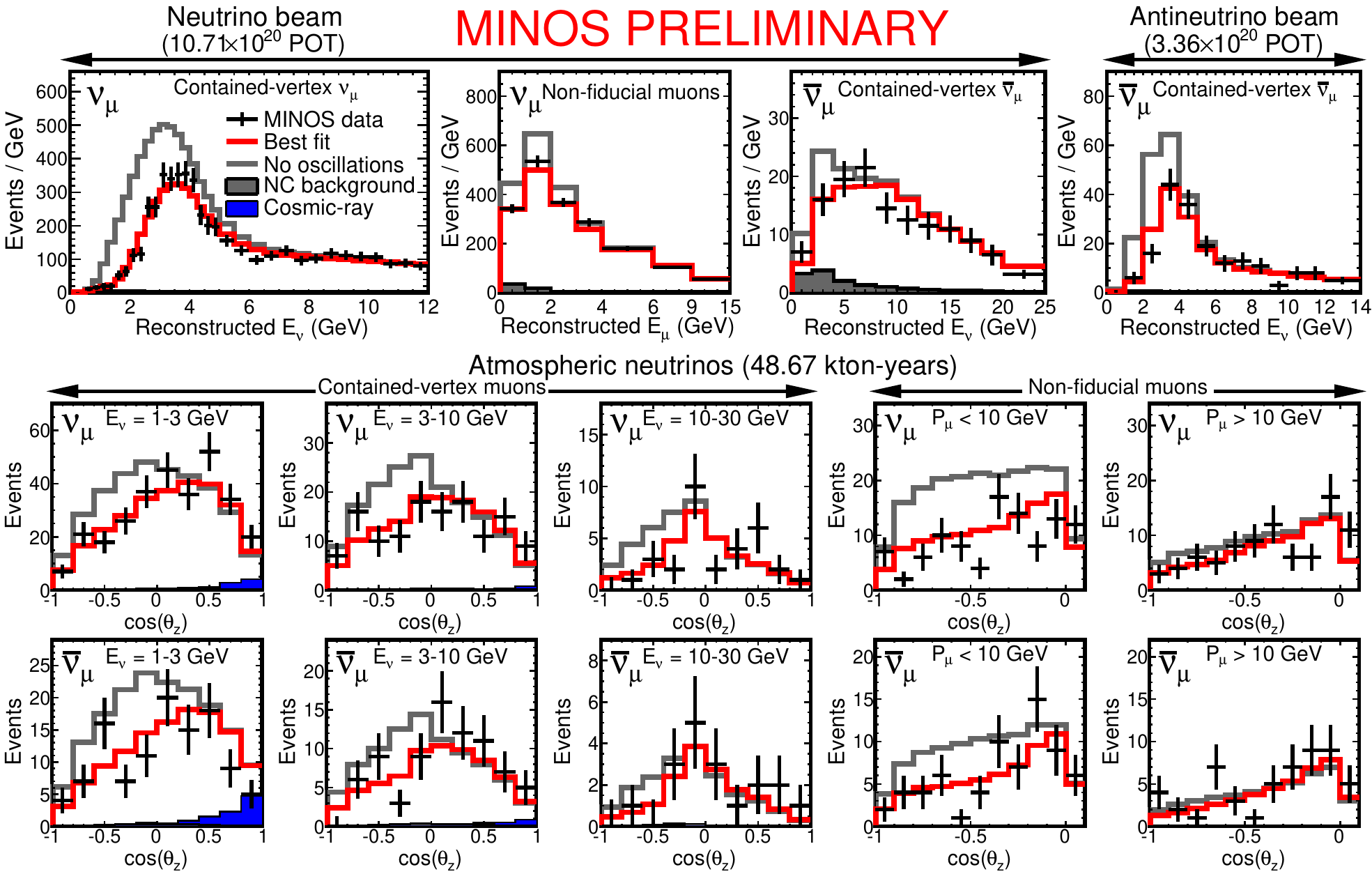}
	\caption{Plots showing the various samples used in the analysis. The top plots show MINOS beam data. The others are MINOS and MINOS+ atmospheric data for contained-vertex and anti-fiducial samples of neutrinos and antineutrinos. The distributions of atmospheric neutrinos are plotted as a function of zenith angle and separated into bins of reconstructed neutrino energy. The observed data (points) are compared with the prediction for no oscillations (grey) and the best fit oscillations (red). The cosmic ray background is shown in blue while the NC background is shown in gray.}
  \label{fig:samples}
\end{figure}

We present new results from a three-flavor combined analysis of  \numu~disappearance and \numu$\rightarrow\nu_e$~appearance, including the full MINOS beam sample and and an atmospherics sample increased by 28\%  with respect to the sample used in the first results of this analysis~\cite{ref:minosthree}. This increase from 37.88\,kton-years to 48.70\,kton-years results from operation of the FD during the NuMI 2012/2013 shutdown and during the first year of MINOS+ beam operation. The full MINOS beam sample used an exposure of $10.71\times10^{20}$\,protons-on-target from \numu-dominated running, and $3.36\times10^{20}$\,protons-on-target collected while running in $\overline{\nu}_\mu$-enhanced mode. The analysis uses CC interactions of both muon and electron neutrinos. These events are distinguished from neutral-current (NC) backgrounds by the presence of a muon track or electromagnetic shower, respectively. The selection of accelerator \numu~CC and $\overline{\nu}_\mu$~CC events is based on a multivariate k-Nearest-Neighbor algorithm using input variables characterizing the topology and energy deposition of muon tracks~\cite{ref:rustem}. The selection of accelerator \nue~CC and  $\overline{\nu}_e$~CC events employs a library-event-matching (LEM) algorithm that performs hit-by-hit comparisons of shower-like events with a large library of simulated neutrino interactions~\cite{ref:LEM}. Finally, atmospheric neutrinos are separated from the cosmic ray backgrounds using selection criteria that identify either a contained-vertex interaction or an upward-going or horizontal muon track~\cite{ref:atmos}. For contained-vertex events, the background is reduced by checking for coincident energy deposits in the veto shield. The event selection yields samples of contained-vertex and non-fiducial muons, which are each separated into candidate \numu~CC and $\overline{\nu}_\mu$~CC interactions. A total of 2579 (312) fiducial \numu~($\overline{\nu}_\mu$~CC) events are observed in the FD, compared to 3201 (363) predicted in the absence of oscillations. The atmospheric sample includes 1134 contained-vertex \numu+$\overline{\nu}_\mu$ events, 590 neutrino-induced muon events, and 899 contained-vertex showers. The reconstructed neutrino energy spectra at the FD for the various samples included in the fit is shown on~Fig.~\ref{fig:samples}. 

The spectra are fitted to the predicted FD spectra (extrapolated from the ND measurement in the beam sample case). Three-flavor neutrino oscillations are applied to the FD predicted spectra using probabilities calculated directly from the PMNS matrix using algorithms optimized for computational efficiency~\cite{ref:oscprob}. The mixing angle $\theta_{13}$ is subject to an external constraint of $\sin^2 \theta_{13} = 0.0242\pm0.0025$, based on a weighted average of the published results from the Daya~Bay, RENO, and Double~Chooz reactor experiments~\cite{ref:dayabay, ref:reno, ref:doublechooz}. The likelihood function contains 32 nuisance parameters that account for the major systematic uncertainties in the simulation of the data~\cite{ref:minosatmos, ref:LEM, ref:minos}. The separate likelihood contributions from the \numu~disappearance and \nue~appearance data sets are summed in the fit, assuming their systematic parameters to be uncorrelated. 

\begin{figure}[h!]
	\centering
  		\includegraphics[width=0.38\textwidth]{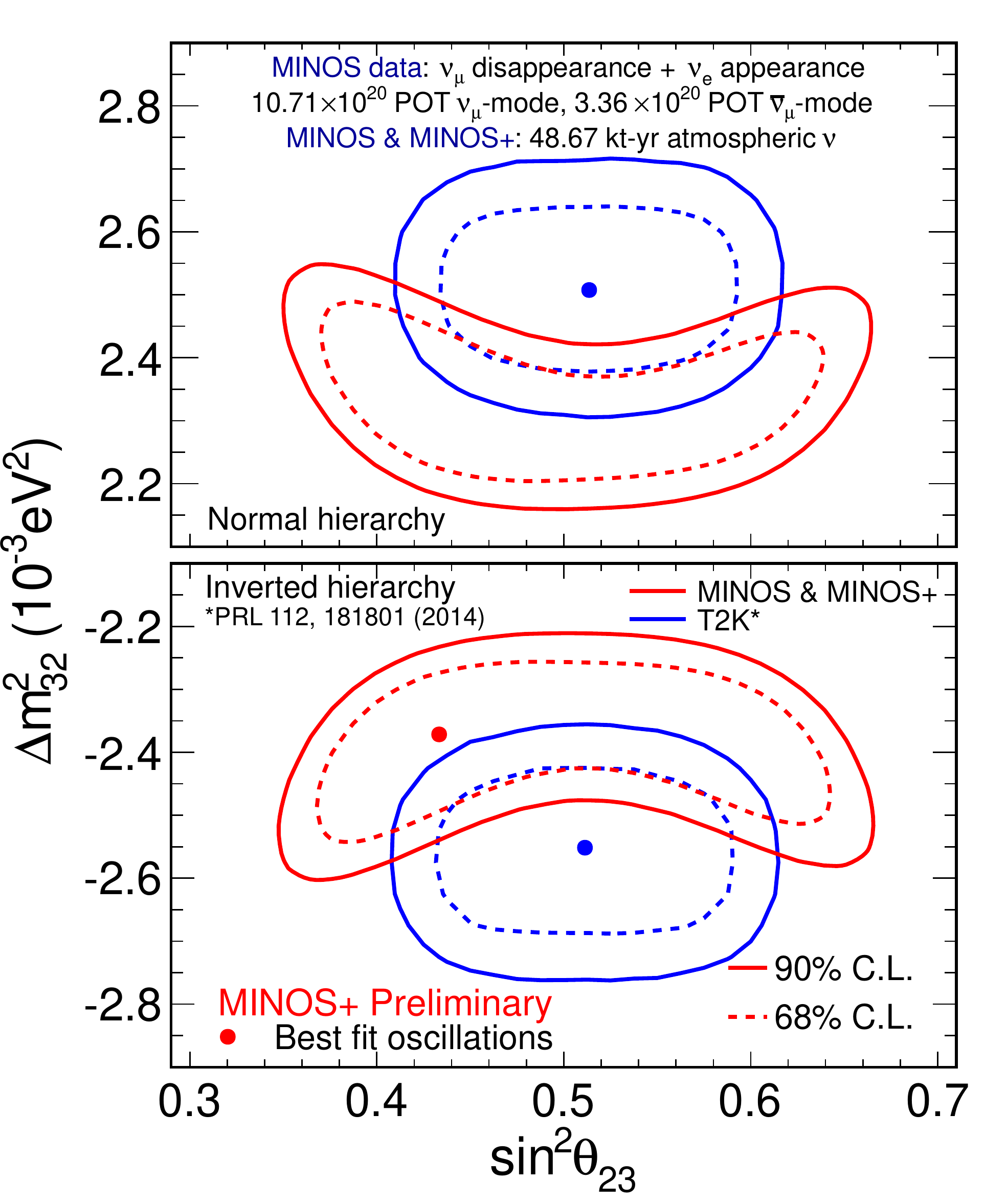}
	\caption{Comparison of ($\Delta m^{2}_{32}$, $\sin^{2}\theta_{23}$) contours from the MINOS three-flavor combined analysis and T2K results~\cite{ref:t2k}.}
 \label{fig:minosvst2k}
\end{figure}

The analysis is sensitive to the value of $\theta_{13}$, the octant of $\theta_{23}$, the neutrino mass hierarchy, and the value of $\delta_{CP}$. It is the first analysis sensitive to the MSW~\cite{ref:MSW} resonance predicted to occur in multi-GeV, upward-going atmospheric neutrinos~\cite{ref:petcov}, by separating atmospheric \numu~CC and $\overline{\nu}_\mu$~CC interactions which will have different yields depending on the octant of $\theta_{23}$ and the neutrino mass hierarchy. The best fit values of the oscillation parameters obtained for the combined fit are $|\Delta m^2_{32} |=2.37^{+0.11}_{-0.07}\times 10^{-3}$\,eV$^2$ and $\sin^2\theta_{23}=0.43^{+0.19}_{-0.05}$  for inverted hierarchy, and $|\Delta m^2_{32} |=2.34^{+0.09}_{-0.09}\times 10^{-3}$\,eV$^2$ and $\sin^2\theta_{23}=0.43^{+0.16}_{-0.04}$ for normal hierarchy. The 90\% confidence limit (C.L.) on the $\theta_{23}$ mixing angle is  $\sin^2\theta_{23}=0.36-0.65$ (90\% C.L.) for the inverted hierarchy and $\sin^2\theta_{23}=0.37-0.64$ for normal hierarchy. The 68\% (90\%) C.L. is calculated by taking the range of negative log-likelihood values with $-2\Delta\log\mathcal{L} < 1.00\,(2.71)$ relative to the overall best fit. The fit has a marginal preference for inverted hierarchy ($-2\Delta\log\mathcal{L} =0.16$), the lower octant of $\theta_{23}$ ($-2\Delta\log\mathcal{L} =0.38$), and non-maximal mixing ($-2\Delta\log\mathcal{L} =0.66$). The 68\% and 90\% C.L. allowed regions in oscillation parameter space are presented in Fig.~\ref{fig:minosvst2k}, compared to the results published by the T2K Collaboration~\cite{ref:t2k}. These are the most precise measurements of $\Delta m^2_{32}$ reported to date. 
\vspace{-4pt}
\vspace{-25pt}

\subsection{First Look at MINOS+ Beam Data and MINOS+ Prospects}
\begin{figure}[h!]
	\centering
	 	\includegraphics[width=0.33\textwidth]{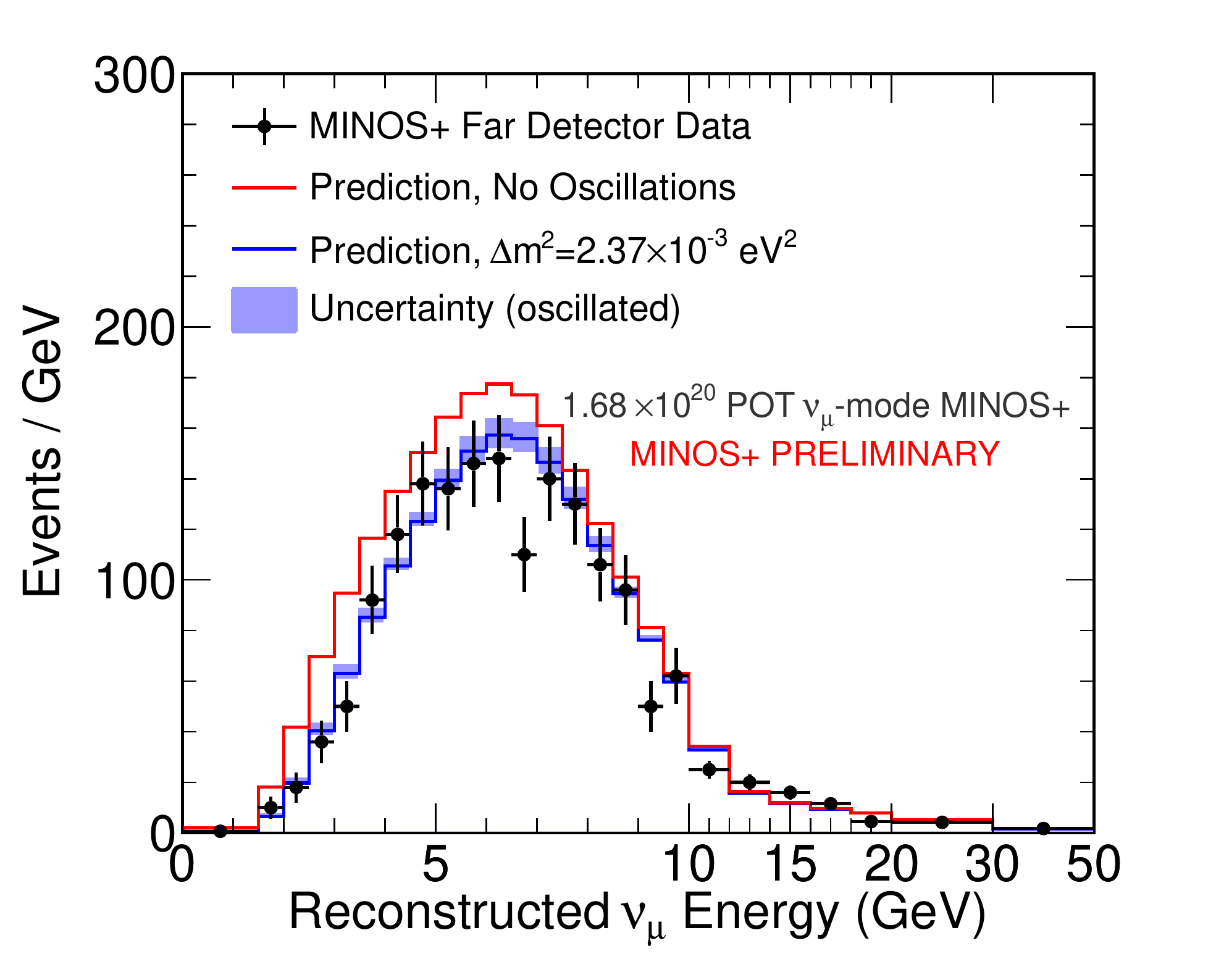}
 	 	\includegraphics[width=0.33\textwidth]{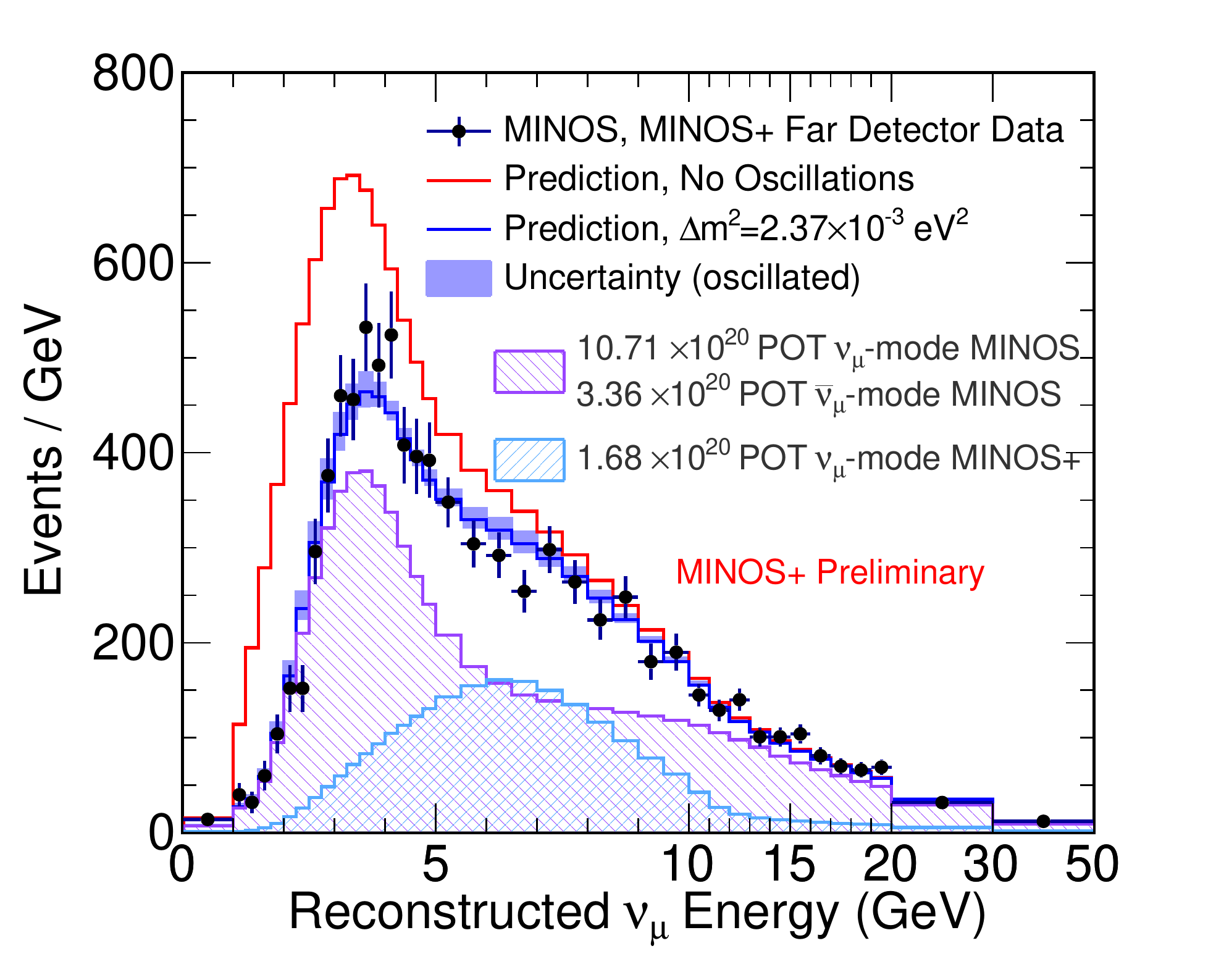}
		\includegraphics[width=0.33\textwidth]{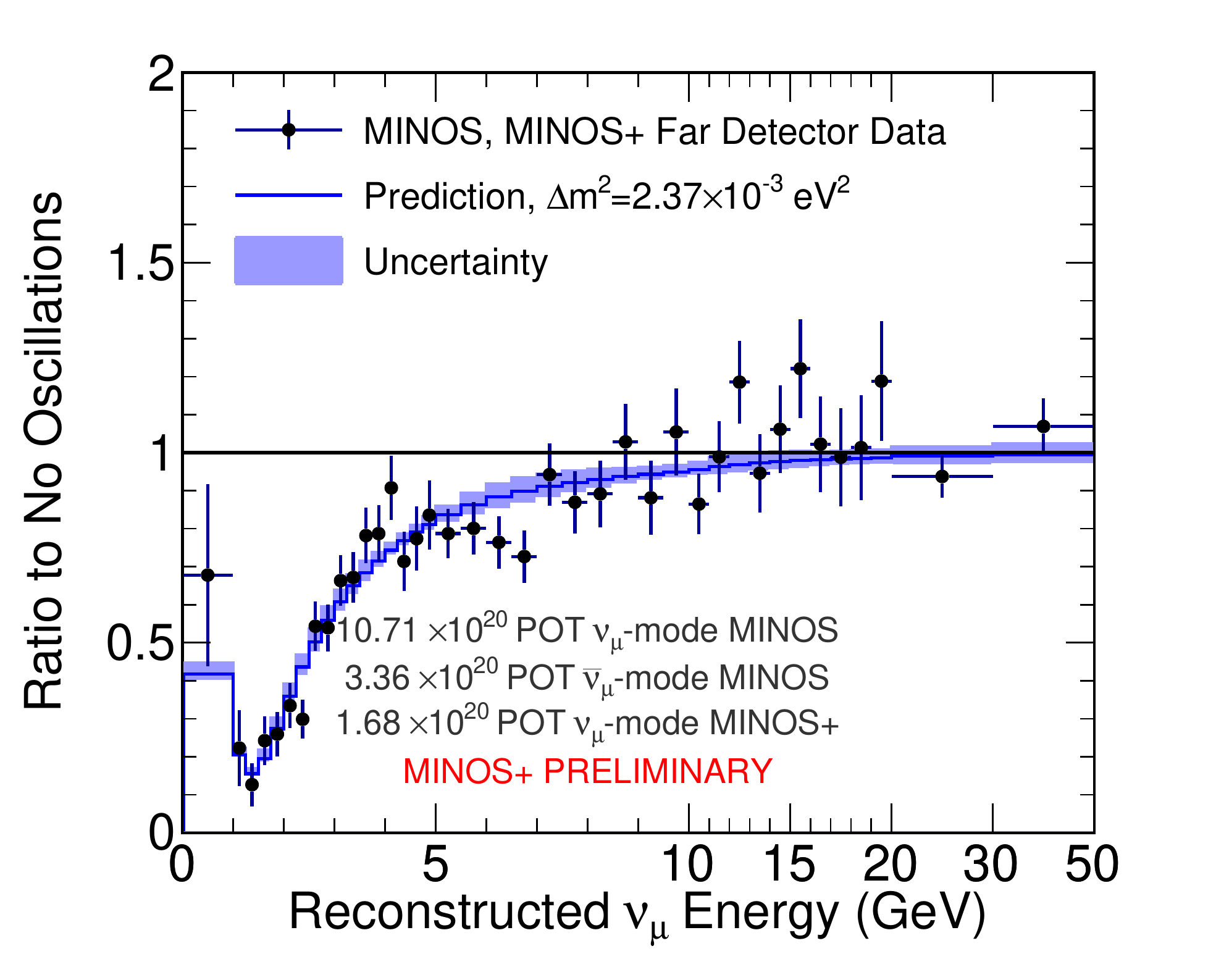}
	\caption{The plot on the left shows the reconstructed energy spectrum for selected FD \numu~CC events in the MINOS+ data (black dots) and prediction at the MINOS best fit oscillations (blue histogram). The center panel shows the result of adding together the MINOS and MINOS+ spectra. The dashed histograms show the predicted contributions of MINOS (purple) and MINOS+ (cyan) assuming MINOS best fit oscillations. The plot on the right displays the oscillated/unoscillated ratio of the combined spectra.}
  \label{fig:minospluscombo}
\end{figure}

\vspace{-8pt}
From September 4, 2013 to April 24, 2014,  MINOS+ collected $1.68\times10^{20}$\,protons-on-target with NuMI running in \numu~mode. Applying the same selection algorithms used in MINOS, but optimized for the medium-energy tune of the NuMI beam, 1037 \numu~CC and 48 $\overline{\nu}_\mu$~CC candidates are found in the FD, compared to an unoscillated prediction of 1255 and 52 events, respectively. The MINOS+ data is consistent with the neutrino oscillation parameters measured by MINOS, as shown in Fig.~\ref{fig:minospluscombo}.
\vspace{-5pt}

\begin{figure}[h!]
	\centering
		\includegraphics[width=0.33\textwidth]{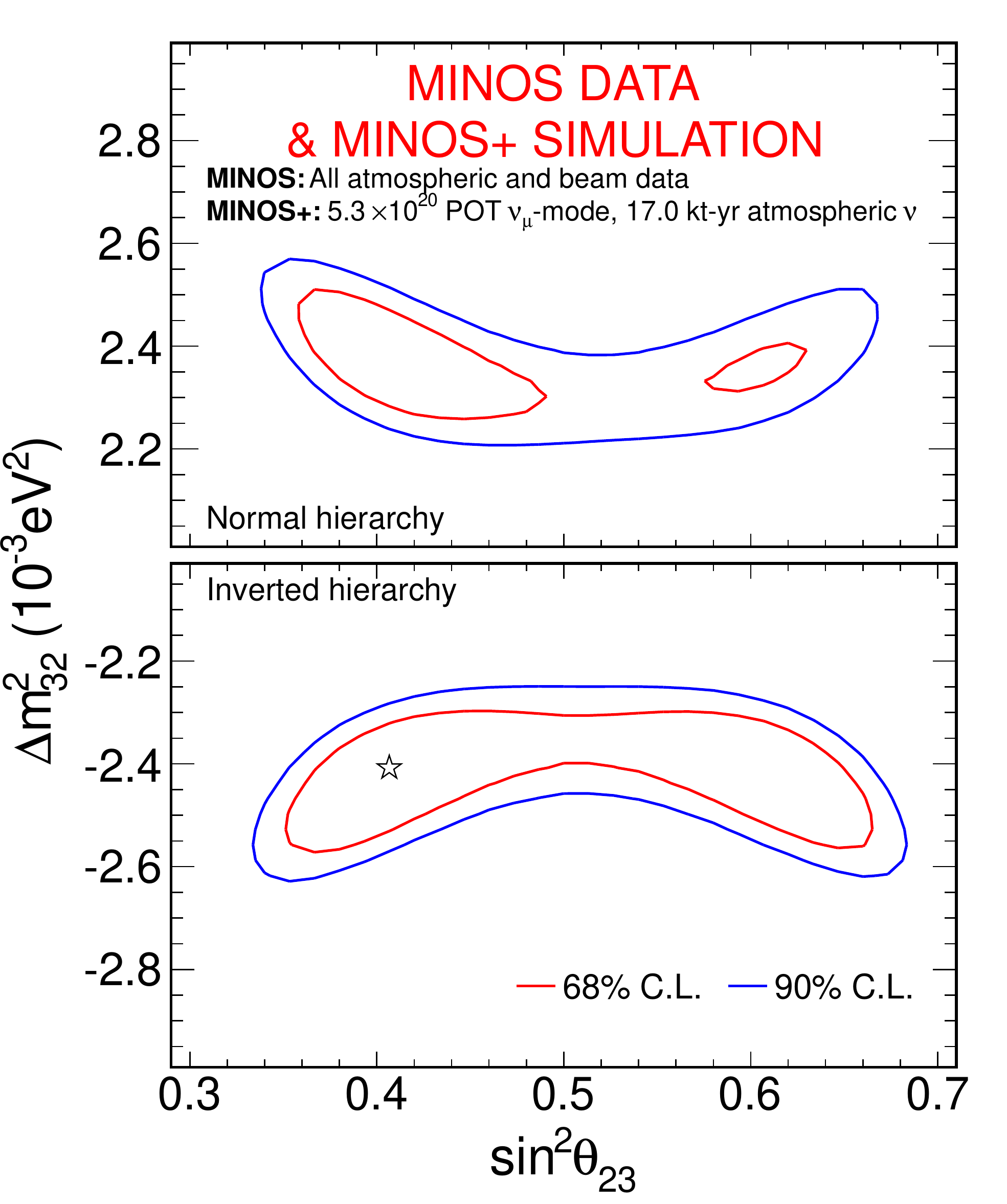}	
 	 	\includegraphics[width=0.33\textwidth]{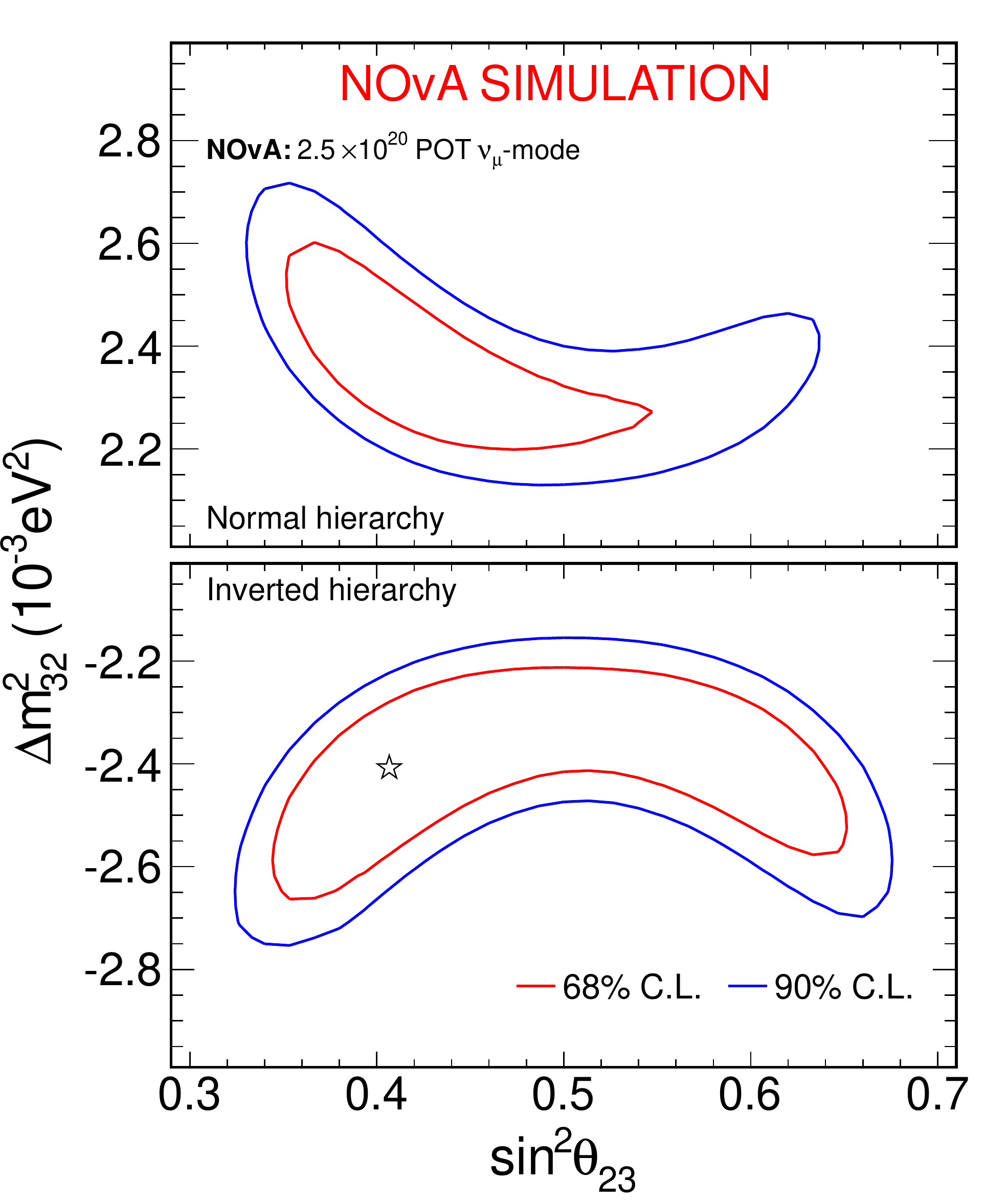}
		\includegraphics[width=0.33\textwidth]{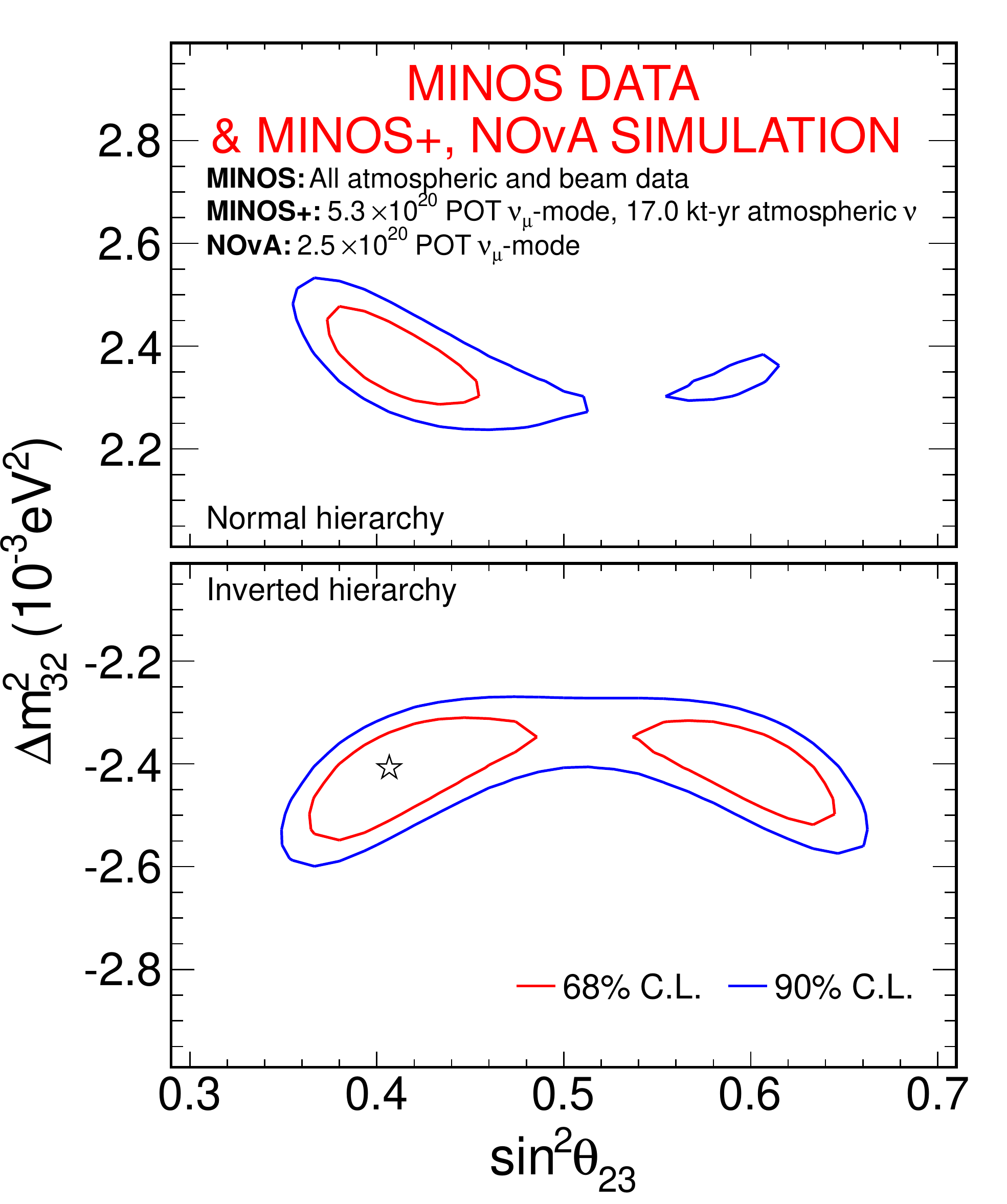}
	\caption{Oscillation parameter sensitivities for predicted exposures by the 2015 Fermilab accelerator shutdown assuming MINOS best fit oscillations. MINOS+-only is show on the left, NO$\upnu$A-only in the center, and a combination of MINOS data, MINOS+ simulation, and NO$\upnu$A simulation on the right. The combined result almost fully excludes the [higher $\theta_{23}$ octant - normal hierarchy] solution at 90\% C.L.}
  \label{fig:minosplusfuture}
\end{figure}

The incorporation of MINOS+ beam data in the fits discussed in the previous section is ongoing and results are expected in 2015 from the analysis of a large statistics exposure. The addition of the MINOS and MINOS+ reconstructed energy spectra results in the best resolution yet of the $P(\numu\rightarrow\numu)$ survival probability curve. The summed spectrum and oscillated/unoscillated ratio are shown on Fig.~\ref{fig:minospluscombo}.   
At the time of writing of these proceedings, MINOS+ has already accumulated $\sim3\times10^{20}$\,protons-on-target. By the 2015 accelerator shutdown at Fermilab, scheduled for June 2015, we expect a MINOS+ exposure of $5.3\times10^{20}$\,protons-on-target. By the same time, NO$\upnu$A will have accumulated $\sim2.5\times10^{20}$\,protons-on-target. Therefore, during NO$\upnu$A's ramp-up, a combination of MINOS+ and NO$\upnu$A will provide the best precision on \numu~disappearance measurements, as shown in Fig.~\ref{fig:minosplusfuture}. As NO$\upnu$A, T2K, and Daya Bay will increase their exposure and precision on the measurement of $\Delta m^2$, the focus of MINOS+ will shift to tests of the three-flavor mixing paradigm by looking for exotic phenomena.

\vspace{-20pt}
\section{Search for Non-Standard Neutrino Interactions in MINOS} 
\begin{figure}[h!]
	\centering
		\includegraphics[width=0.31\textwidth]{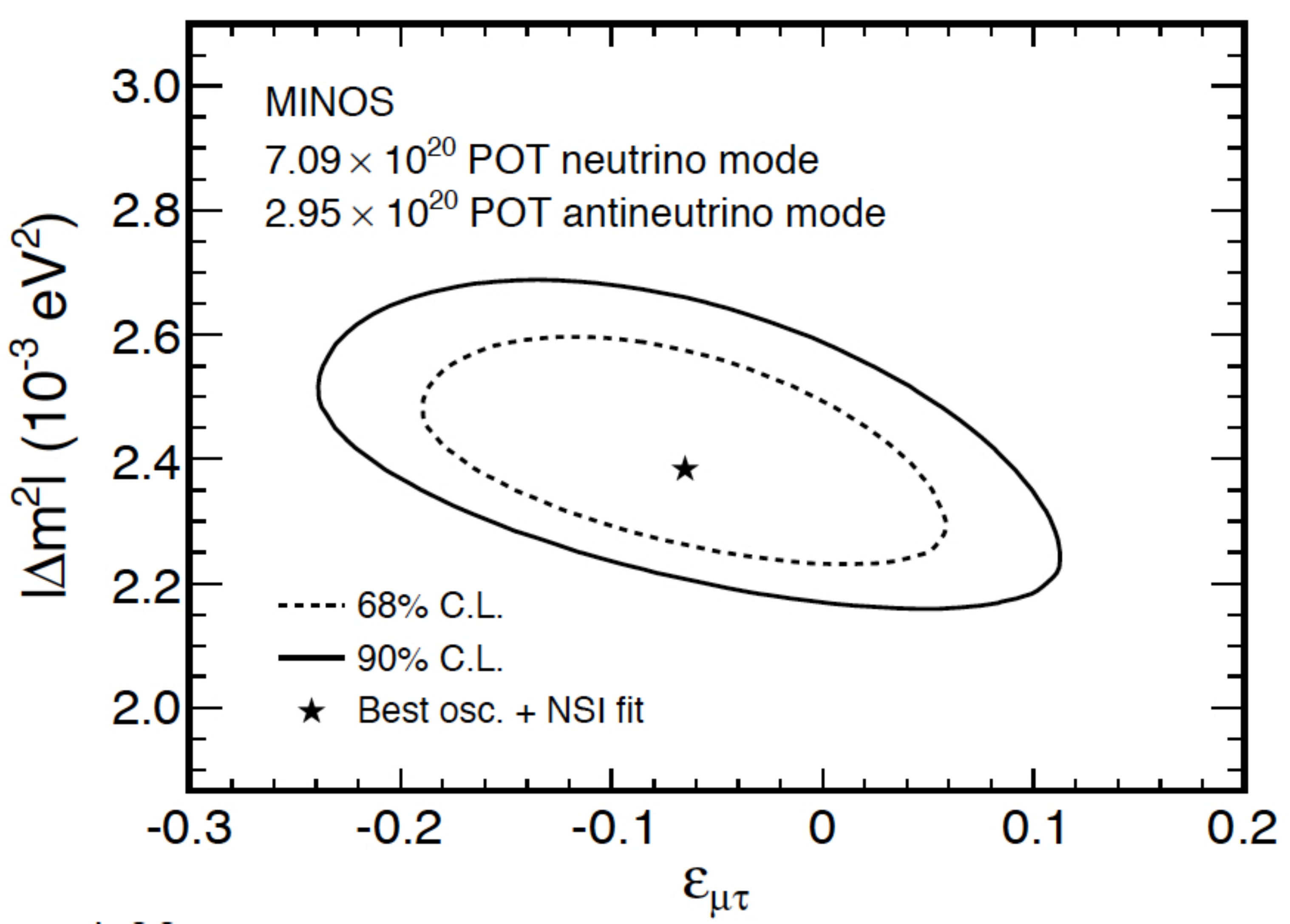}
 	 	\includegraphics[width=0.38\textwidth]{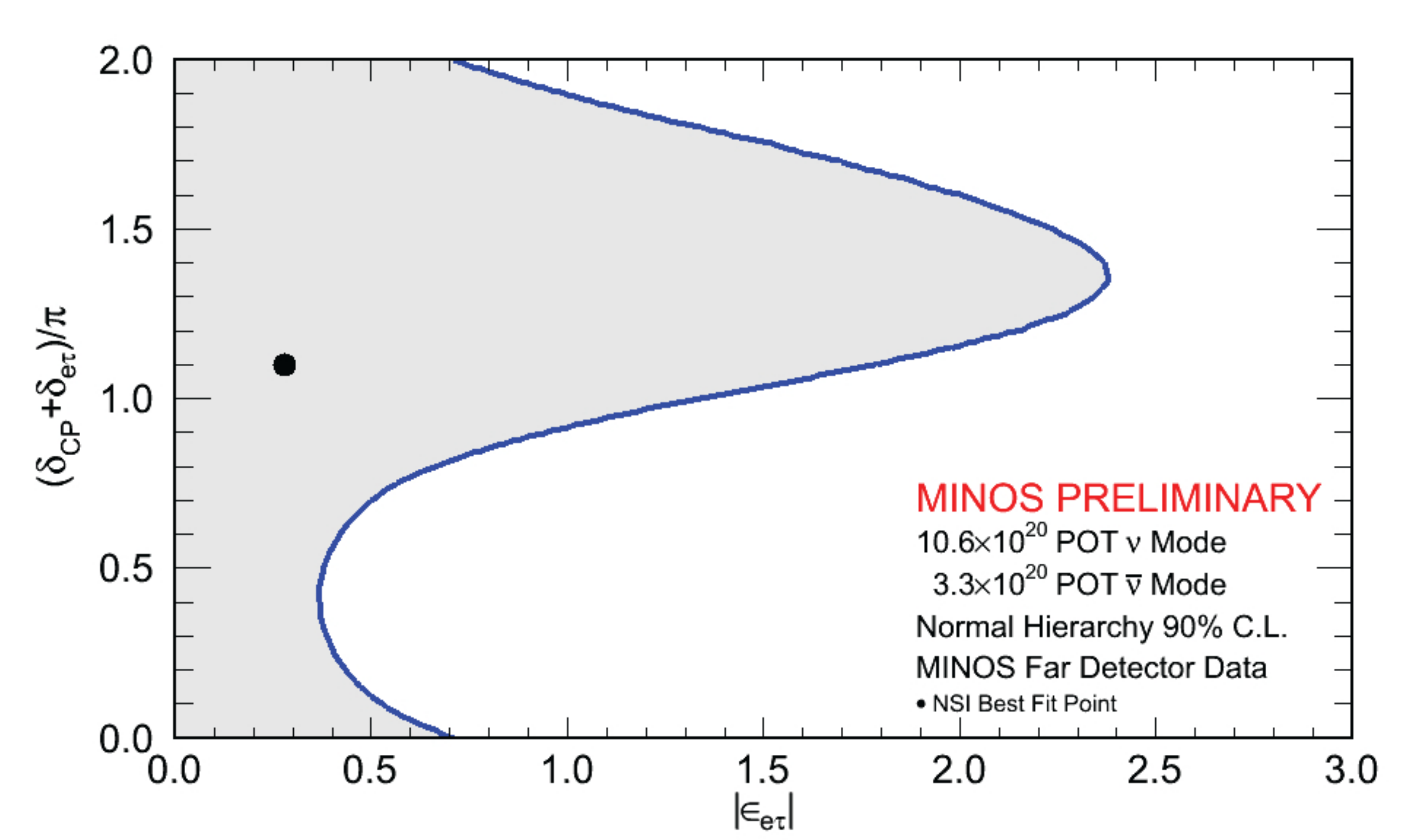}
		\includegraphics[width=0.38\textwidth]{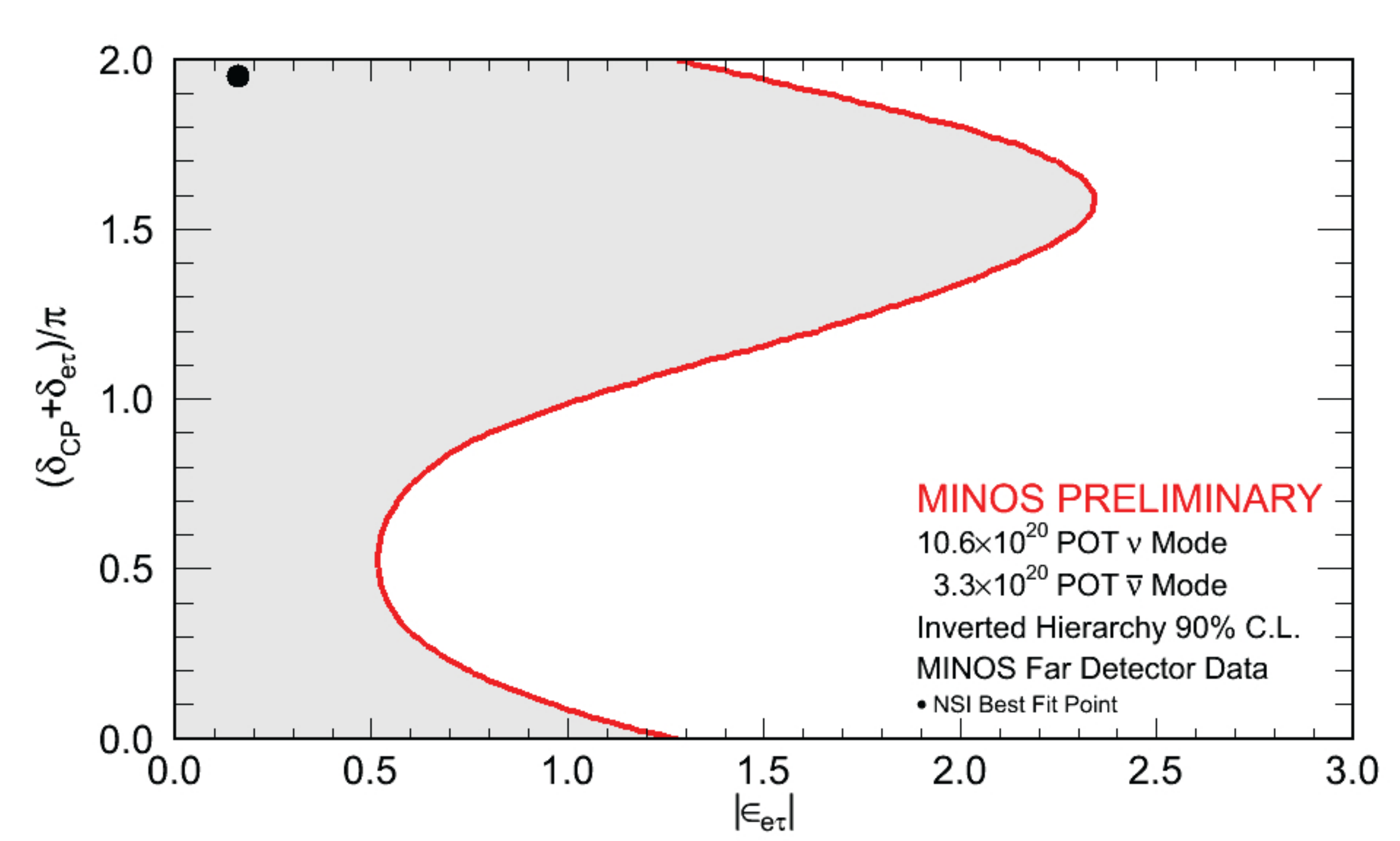}
	\caption{The 68\% and 90\% C.L. allowed regions for $\epsilon_{\mu\tau}$ using MINOS \numu~and $\overline{\nu}_\mu$~disappearance samples (left). Results from a new \nue~appearance-based search for NSI. The center plot shows the allowed region and best fit for $\epsilon_{e\tau}$ in shaded gray if normal hierarchy is assumed; the plot on the right shows the same allowed region for inverted hierarchy. In both cases, the contours are marginalized over $\delta_{CP}$.}
	\label{fig:minosnsi}
\end{figure}

Non-standard neutrino interactions (NSI) are a generic modification of the neutrino hamiltonian that can alter the flavor content of a neutrino beam as it propagates through the Earth's crust, in a manner akin to standard matter effects~\cite{ref:nsi}. MINOS has previously conducted searches for neutral-current NSI using \numu~and $\overline{\nu}_\mu$~disappearance, placing limits on the $\epsilon_{\mu\tau}$ coefficient: $-0.20<\epsilon_{\mu\tau}<0.07$ (90\% C.L.)~\cite{ref:minosnsi}. We present results from a new NSI search using the complete MINOS \nue~appearance sample and the same \nue~CC selection methodology described above. This search follows the prescriptions in Refs.~\cite{ref:nsipheno, ref:nsitony} and is sensitive to the $\epsilon_{e\tau}$ NSI coefficient. Results of the analysis are shown in Fig.~\ref{fig:minosnsi} and significantly improve on the global $\epsilon_{e\tau}$ limit $|\epsilon_{e\tau}|<3.0$ reported in Ref.~\cite{ref:nsi_ohlsson}.

\section{Search for Sterile Neutrinos in the MINOS Data} 
Most results from experiments measuring solar neutrinos, atmospheric neutrinos, and neutrinos produced by accelerators and in nuclear reactors, are well-described by oscillations between three distinct neutrino types: the electron, muon, and tau neutrinos.
However, several anomalies have puzzled and baffled the neutrino physics community. For example, the Liquid Scintilator Neutrino Detector (LSND) experiment reported a 3.8\,$\sigma$ excess of $\overline{\nu}_e$ appearance in a $\overline{\nu}_\mu$ beam over a short baseline~\cite{ref:lsnd}. This excess has been interpreted as evidence for oscillations between the known active neutrinos and so-called sterile neutrinos, which do not couple to other matter through known Standard Model forces. Additional neutrino flavors may clarify the origin of neutrino mass, provide dark matter candidates, and explain core collapse in supernov\ae. These strongly compel searches for sterile neutrinos. 

\begin{figure}[!h]
	\vspace{0pt}
	\centering
	  	\includegraphics[width=0.39\textwidth]{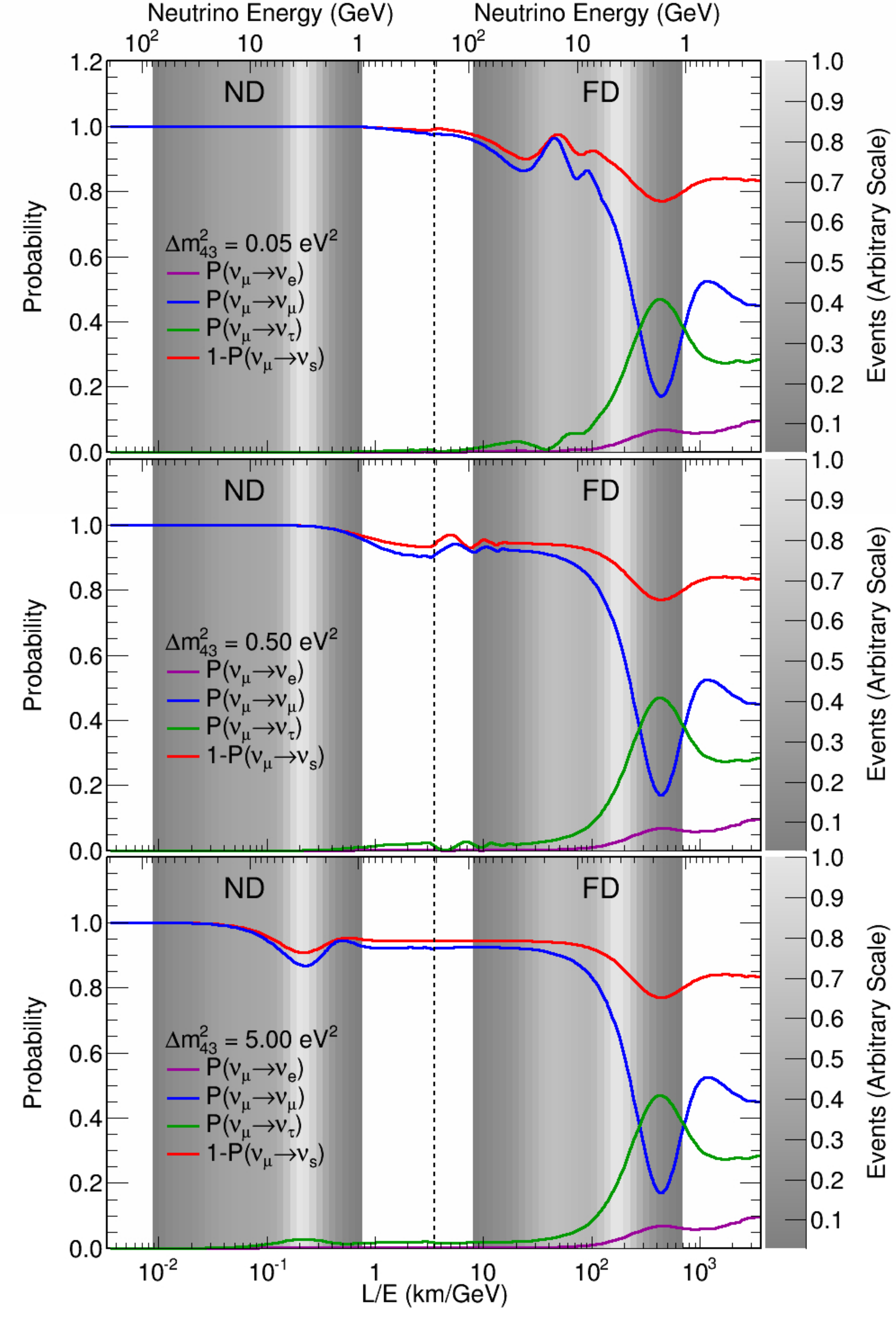}	
\vspace{0pt}
\caption{Regions of $L/E$ probed by the MINOS+ detectors compared to neutrino disappearance and appearance probabilities. The scale at the top shows the neutrino energy measured by the detectors, increasing from right to left. The top plot shows the probabilities assuming mixing with one sterile neutrino with \delm$_{\rm{43}}=0.05$~eV$^2$, corresponding to the Slow oscillations regime. In this case the ND is unaffected but the FD is sensitive to distortions in the three-flavor \numu~survival probability (blue curve) for neutrino energies $\sim15$\,GeV, well above the minimum of $1.6$\,GeV at 735\,km. The middle plot assumes mixing with one sterile neutrino with \delm$_{\rm{43}}=0.5$~eV$^2$, corresponding to the Intermediate oscillations regime. The ND is insensitive to oscillations, but the oscillations are rapid enough that they average out at the FD. The bottom plot includes mixing with one sterile neutrino with \delm$_{\rm{43}}=5$~eV$^2$, corresponding to the Rapid oscillations regime. For this case, oscillations affect the ND measurement.}
  \label{fig:regimes}
 \vspace{0pt}
\end{figure}

MINOS searches for sterile neutrinos through disappearance measurements, by looking for depletion both in NC interactions, which are topologically independent of neutrino flavor and therefore insensitive to standard three-flavor oscillations, and also in \numu~CC interactions, where mixing with a sterile neutrino would be evident in deviations from the three-active-neutrino mixing pattern. MINOS has unique capabilities among existing experiments in searching for sterile neutrinos. For instance, for neutrino oscillation scenarios including three active neutrinos and one sterile neutrino, a new independent neutrino mass-squared difference, \delm$_{43}$, is required, and three new mixing angles, $\theta_{14}$, $\theta_{24}$, $\theta_{34}$ describe the mixing between active and sterile states. Due to its two detectors separated by 735\,km and beam spectrum peak energy at 3 GeV, MINOS is sensitive to disappearance of \numu~CC and NC interactions over three distinct sterile mixing regimes corresponding to different values of the sterile mass splitting \delm$_{43}$:  1) Slow oscillations (\delm$_{\rm{43}}\lesssim0.1$~eV$^2$); 2) Intermediate oscillations ($0.1\lesssim$\,\delm$_{\rm{43}}\lesssim1$~eV$^2$); and 3) Rapid oscillations (\delm$_{\rm{43}}\gtrsim1$~eV$^2$). The sensitivity to light sterile neutrino masses in MINOS and MINOS+ as a function of \delm$_{43}$ and $L/E$ is illustrated in Figure~\ref{fig:regimes}.

\begin{figure}[!h]
  		\includegraphics[width=0.48\textwidth]{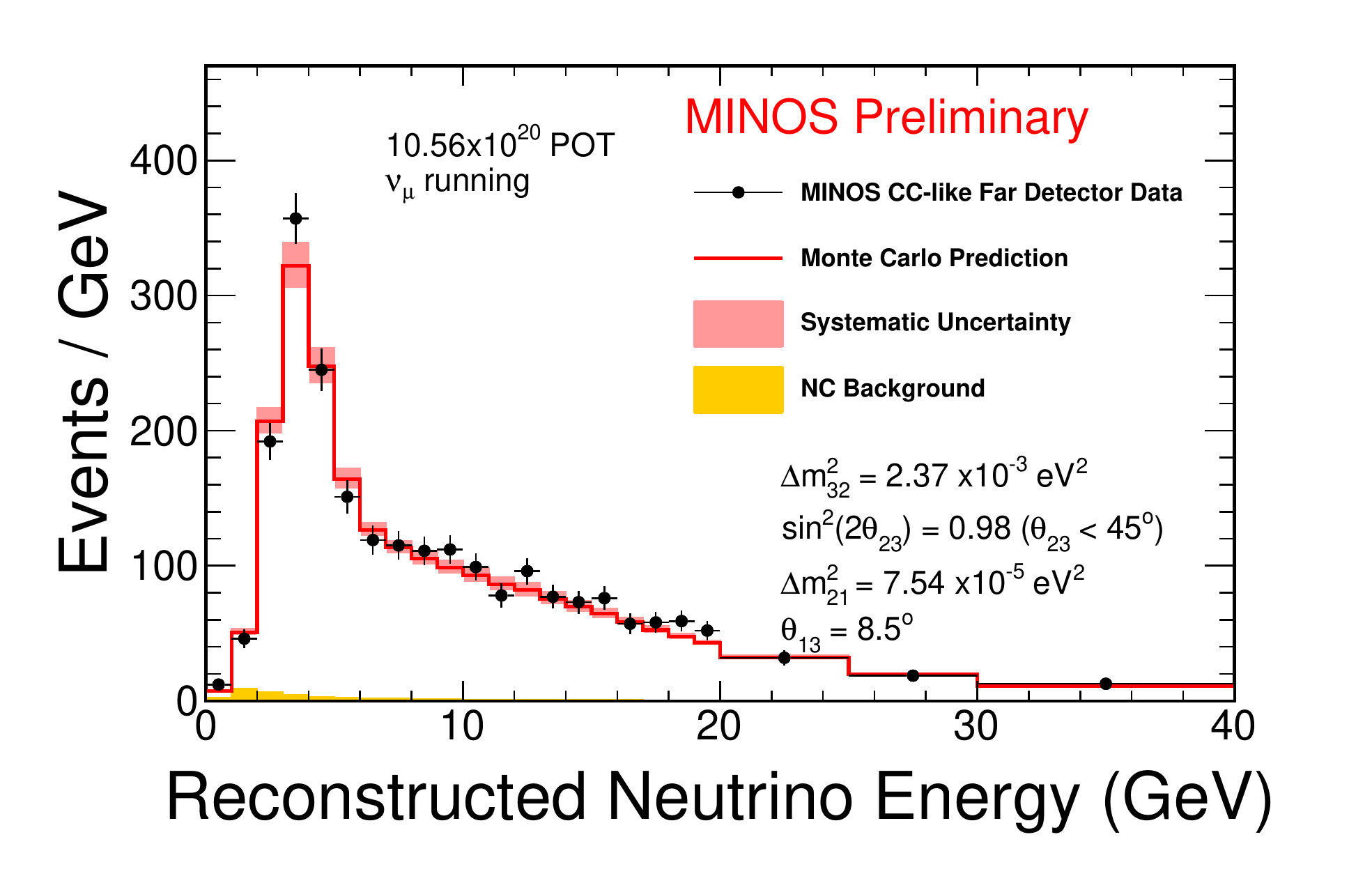}
		\includegraphics[width=0.48\textwidth]{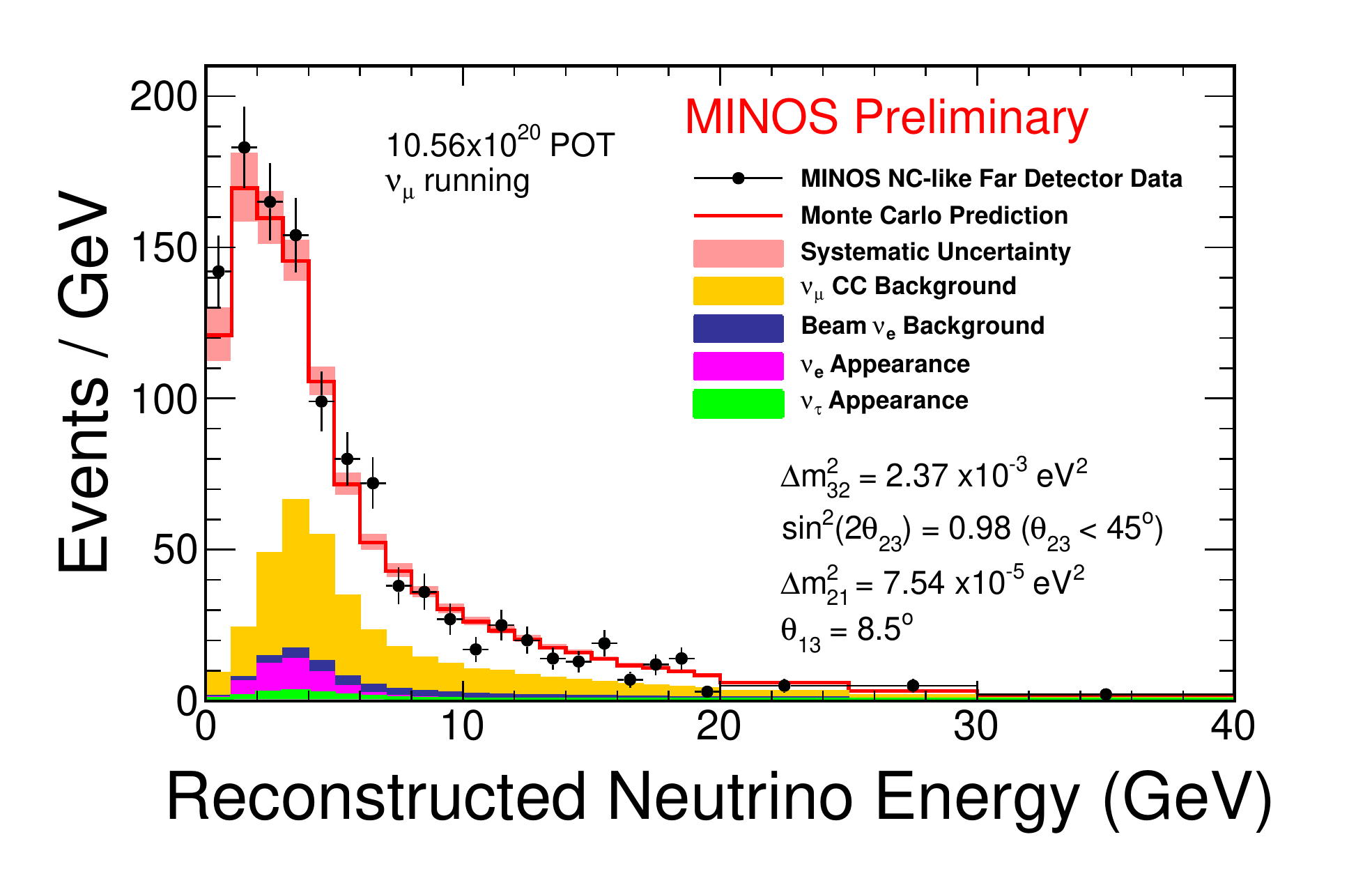}
\caption{Reconstructed FD CC spectrum (left) and FD NC spectrum (right). In both cases the predicted spectrum (red line) is oscillated with the MINOS three-flavor best fit oscillation values. The effect of systematic uncertainties is illustrated by the red band. The backgrounds for each selection are depicted as solid histograms.}
\label{fig:spectra}
\end{figure}

MINOS has excellent reach in probing disappearance into sterile neutrinos, primarily driven by $\theta_{24}$, over 4 orders of magnitude in \delm$_{43}$. This broad reach is particularly relevant for values of \delm$_{43}\lesssim$0.1\,eV$^2$. Previous experiments have not probed for the possibility of \numu~disappearance into sterile neutrinos in this region of parameter space. This sensitivity is enhanced for MINOS+ thanks to large statistics at higher neutrino energies. 
The selection of NC candidates for the analysis uses topological variables related to event length and the relative span of a track compared to the size of a shower to separate \numu~CC backgrounds. This selection method accepts \nue~CC events with high efficiency, so the same external constraint on the mixing angle $\theta_{13}$ described above is applied in the analysis. The events failing the NC selection are then evaluated by the k-Nearest-Neighbor CC selection algorithm. The obtained FD CC and NC reconstructed energy spectra are show in Fig.~\ref{fig:spectra}. Neither the CC or NC spectra show evidence for a depletion consistent with sterile neutrino admixture.
\vspace {-3pt}
\vspace {-0pt}

\begin{figure}[!h]
  		\includegraphics[width=0.48\textwidth]{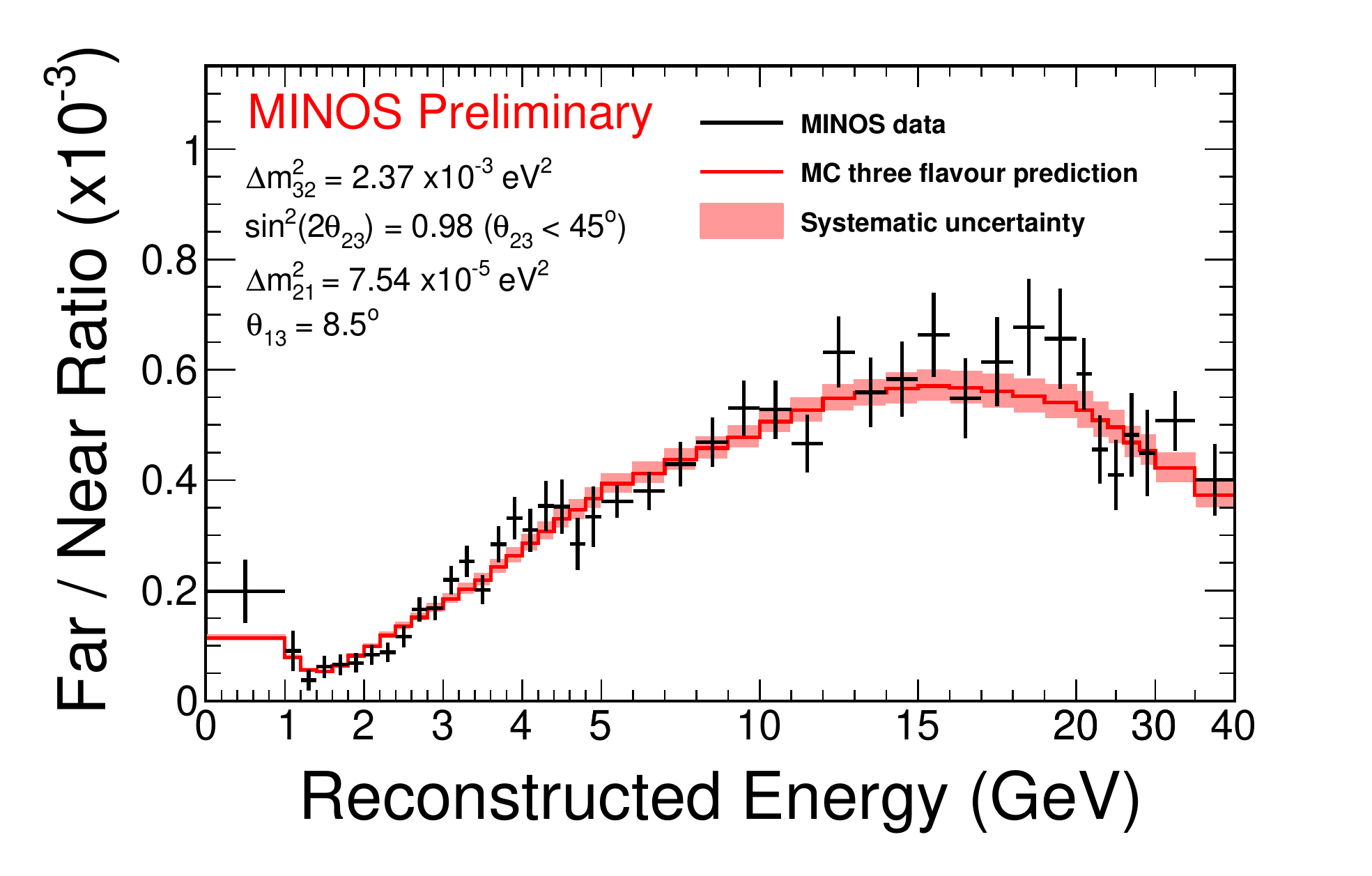}
		\includegraphics[width=0.48\textwidth]{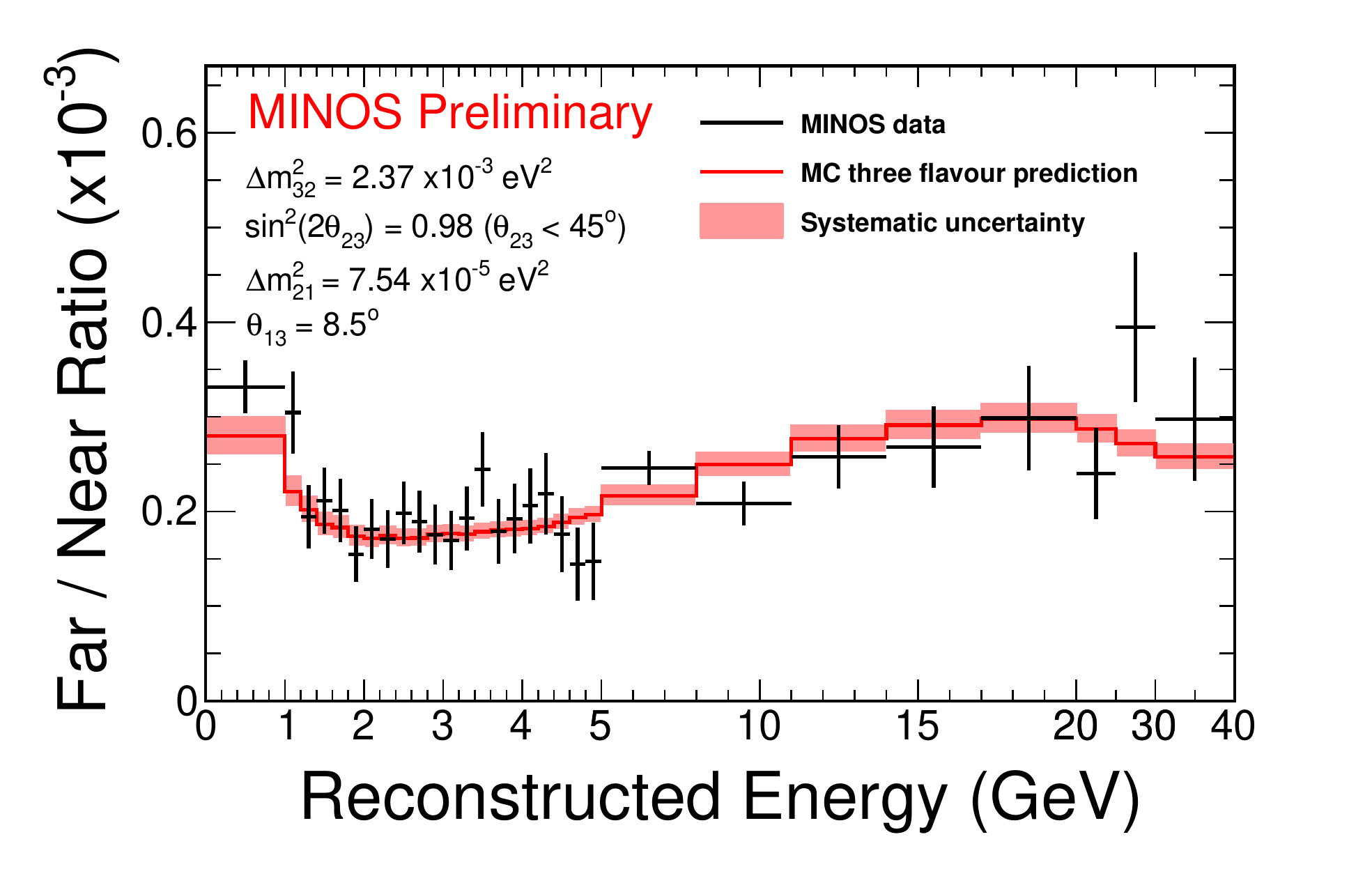}
\caption{Data and Monte Carlo F/N ratios for CC (left) and NC (right) selected events.}
\label{fig:fnratio}
\end{figure}

The four-flavor sterile neutrino analysis underwent several improvements over the previously published sterile neutrino search results~\cite{ref:minossterile}. 3+1 sterile oscillations are now applied to both the ND and FD predictions, and the distance from the neutrino vertex in the ND to the meson decay point is taken into account when computing the oscillations. To account for ND distortions, we fit the oscillated Far/Near Monte Carlo ratio to the Far/Near data ratio, and a penalty term on the ND event rate is included in the fit. The oscillation fit parameters are $|\Delta m^2_{32}|$, $|\Delta m^2_{43}|$, $\theta_{23}$, $\theta_{24}$, and $\theta_{34}$. The CC and NC F/N ratios used in the fit are shown in Fig.~\ref{fig:fnratio}.
\vspace {-4pt}
A thorough reassessment of the systematic uncertainties affecting the higher-energy portion of the spectrum was carried out, including a re-evaluation of beam flux uncertainties. A total of 26 systematic uncertainties are included in the fit via covariance matrices. The log-likelihood surfaces obtained are then corrected for non-gaussianity following the Feldman-Cousins unified approach~\cite{ref:fc}. 

\begin{figure}[!h]
  		\includegraphics[width=0.55\textwidth]{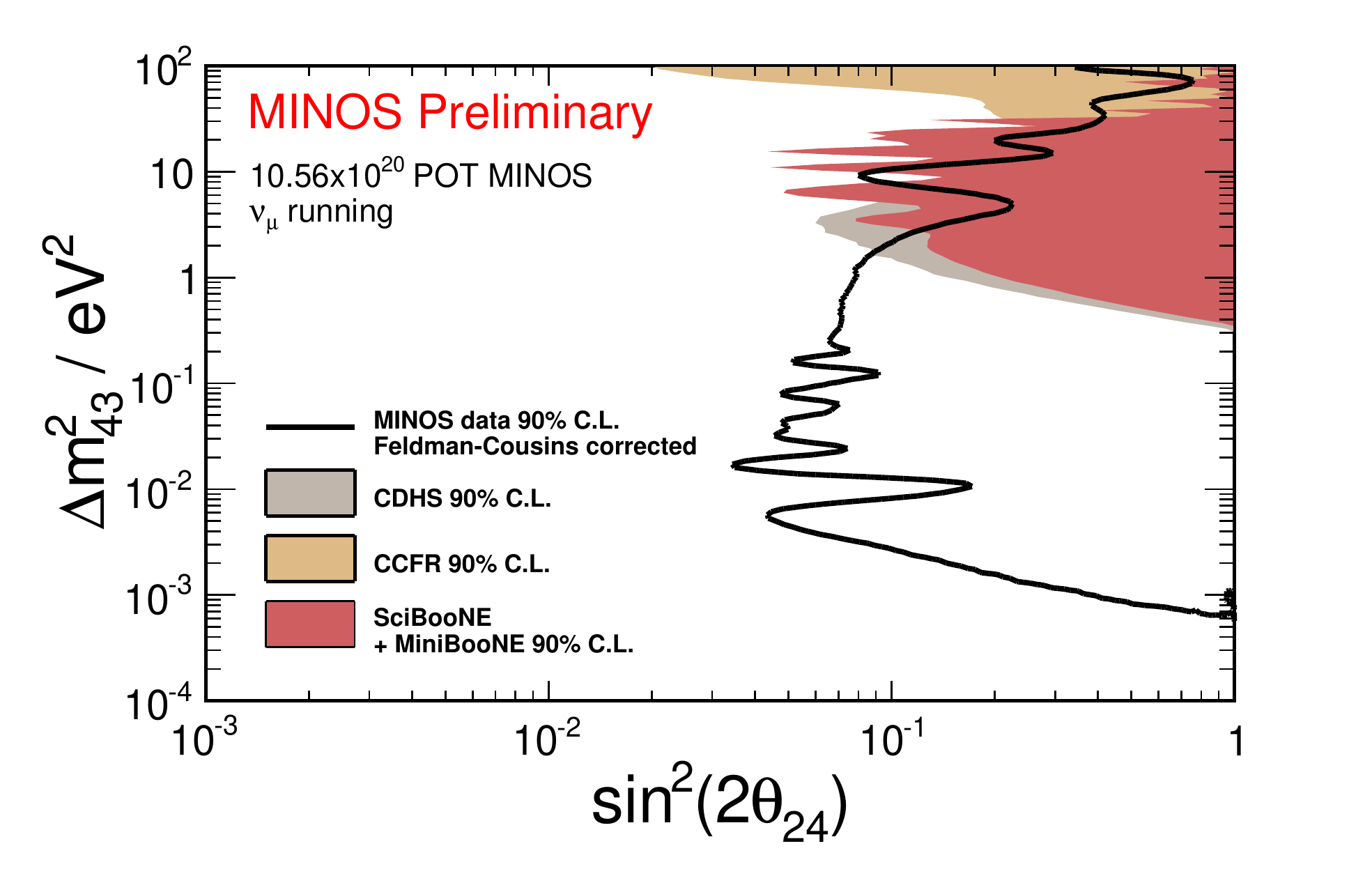}
		\includegraphics[width=0.49\textwidth]{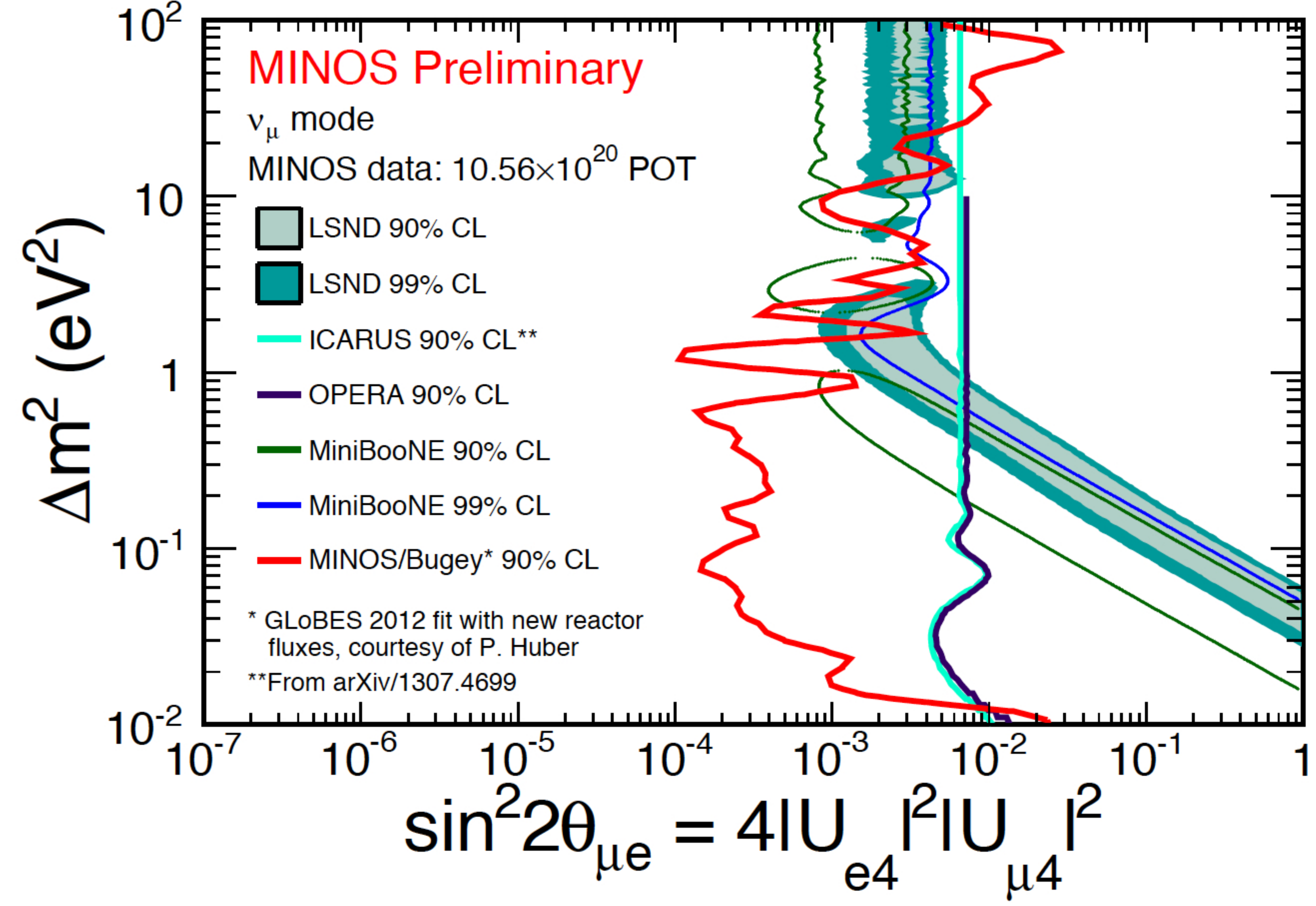}
		
\caption{The left-hand side plot displays the 90\% C.L. disappearance exclusion obtained on the $\theta_{24}$ sterile mixing angle. The right-hand side plot shows results of combining the MINOS limit with the Bugey 90\% C.L. limit on $\nu_e$~disappearance~\cite{ref:bugey} compared to the LSND~\cite{ref:lsnd} and MiniBooNE~\cite{ref:miniboone} allowed regions, and the ICARUS~\cite{ref:icarus} and OPERA~\cite{ref:opera} 90\% C.L. exclusions.}
\label{fig:th24limit}
\end{figure}

The MINOS 90\% C.L. disappearance exclusion obtained on the $\theta_{24}$ sterile mixing angle is shown on Fig.~\ref{fig:th24limit}. The limit spans over 4 orders of magnitude in $|\Delta m^2_{43}|$ and places the stronger constraints yet on \numu~disappearance into sterile neutrinos for $|\Delta m^2_{43}|\lesssim1$~eV$^2$. In order to compare the MINOS constraints with positive results from appearance experiments like LSND and MiniBooNE~\cite{ref:miniboone}, which are presented as a function of $\sin^22\theta_{\mu e}=\sin^22\theta_{14}\sin^2\theta_{24}$, the MINOS 90\% C.L. exclusion is combined with the 90\% C.L. limit on $\theta_{14}$ excluding $\nu_e$~disappearance from the Bugey reactor experiment~\cite{ref:bugey}. The resulting limit, shown in Fig.~\ref{fig:th24limit}, excludes the LSND~\cite{ref:lsnd} and MiniBooNE~\cite{ref:miniboone} allowed regions at 90\% C.L. for $|\Delta m^2_{43}|<1$~eV$^2$, and will increase the existing tension~\cite{ref:tension} between null and signal results for that range of sterile mass values.

\begin{figure}[!h]
  		\includegraphics[width=0.48\textwidth]{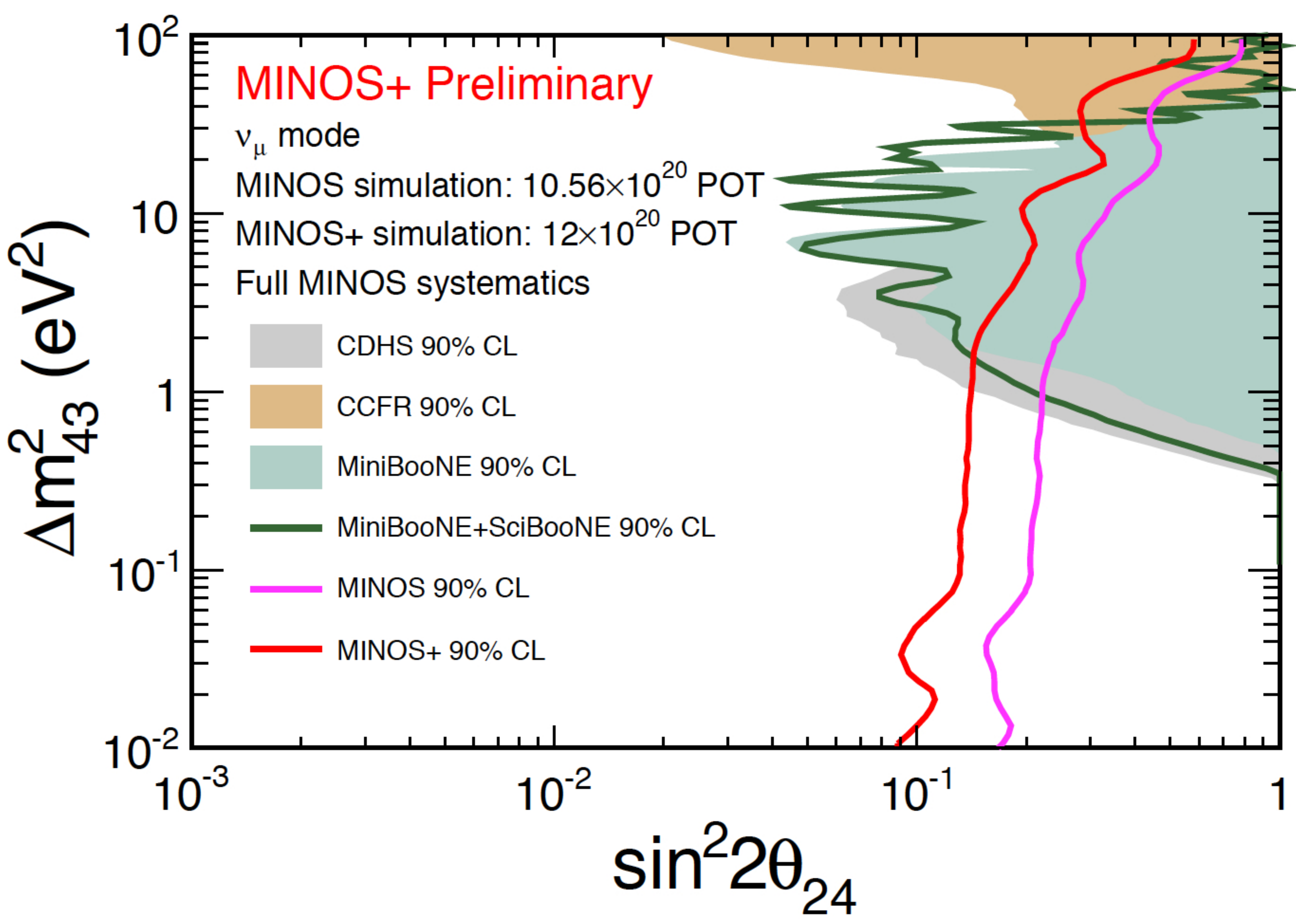}
		\includegraphics[width=0.56\textwidth]{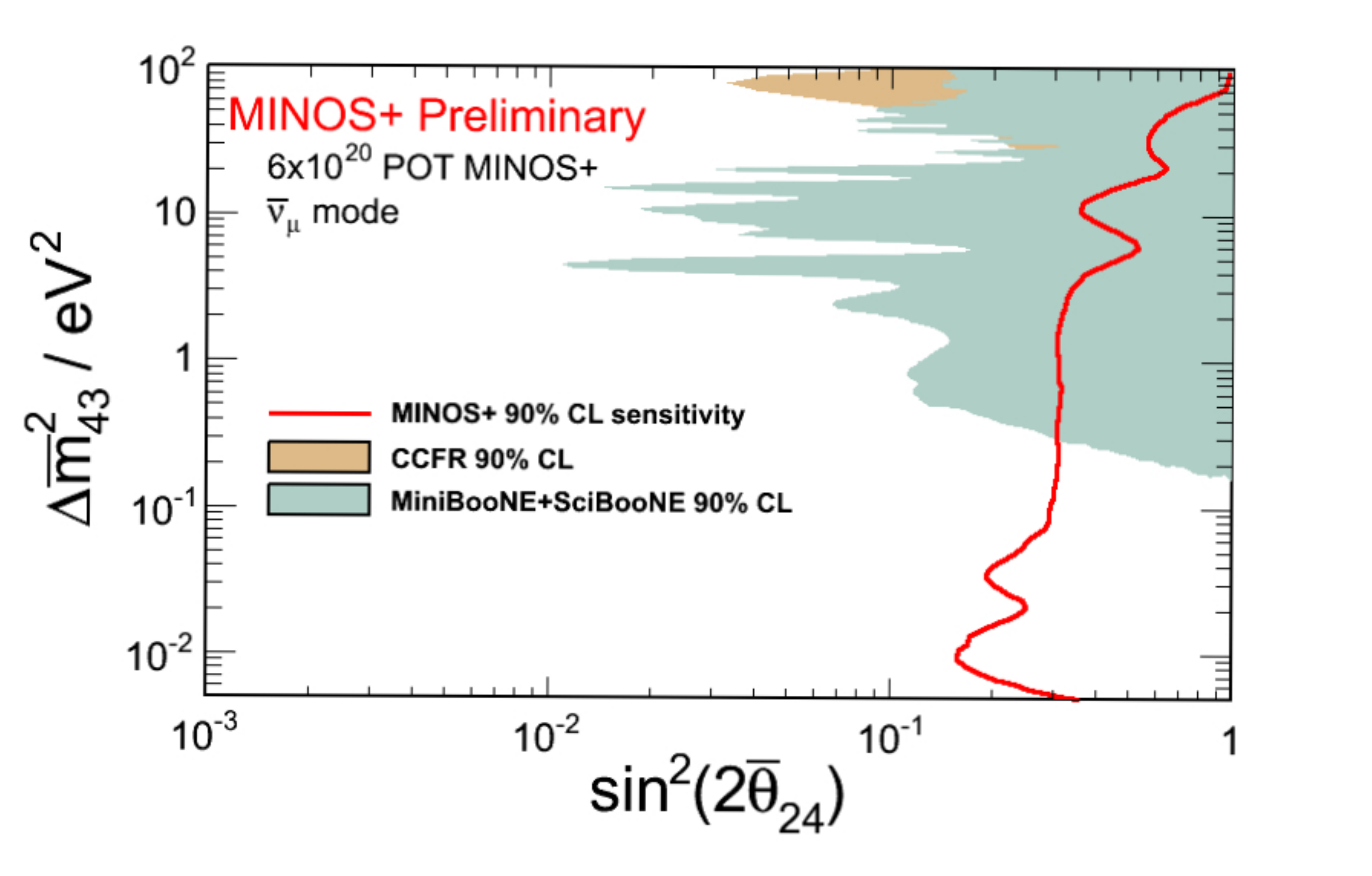}
\caption{MINOS+ reach in probing \numu~disappearance into sterile neutrinos for NuMI neutrino running (left), and NuMI antineutrino running (right). MINOS+ can improve the present MINOS limits by a factor of two for an exposure of $12\times\nobreak10^{20}$~protons-on-target, and probe a new region of parameter space for antineutrino-sterile mixing.}
\label{fig:minosplussterile}
\end{figure}

With large statistics at higher neutrino energies, MINOS+ will improve the MINOS limits and could probe a new region of parameter space for \numu~disappearance if the MINOS+ detectors are in operation during the next NuMI running in antineutrino mode, which will possibly start in the second half of 2016. The MINOS+ sensitivities to disappearance into sterile neutrinos for neutrino and antineutrino running are shown in Fig.\ref{fig:minosplussterile}.

\vspace{-20pt}
\section{Summary}
The MINOS and MINOS+ Collaborations have presented new results from a three-flavor combined disappearance and appearance analysis. The best fit values for oscillation parameters obtained are $|\Delta m^2_{32} |=2.37^{+0.11}_{-0.07}\times 10^{-3}$\,eV$^2$ and $\sin^2\theta_{23}=0.43^{+0.19}_{-0.05}$  for inverted hierarchy. The data are consistent with maximal mixing and have marginal preference for inverted hierarchy and the lower octant of $\theta_{23}$. The beam data from the first year of MINOS+ operation is consistent with the MINOS oscillation measurement. A new search for NSI on MINOS data yields new constraints on the $\epsilon_{e\tau}$ parameters. A new search for sterile neutrinos  places strong constraints on \numu~disappearance into sterile neutrinos for $|\Delta m^2_{43}|\lesssim1$~eV$^2$, and increases the tension between null disappearance and positive appearance results.

\begin{theacknowledgments}
We dedicate the results presented in this work to the memory of U.S. Representative James L. Oberstar (1934-2014), from Minnesota's $8^{\rm{th}}$ Congressional District, for his enthusiastic and unwavering support for the MINOS Project. \\
We thank Patrick Huber, Alexei Smirnov, and Carlo Giunti for useful discussions. \\
This work was supported by the U.S. DOE; the United
Kingdom STFC; the U.S. NSF; the State and University
of Minnesota; Brazil's FAPESP, CNPq and CAPES. We
are grateful to the Minnesota Department of Natural Resources
and the personnel of the Soudan Laboratory and
Fermilab for their contributions to the experiment. 
\end{theacknowledgments}







\end{document}